\documentclass[
reprint,
notitlepage,
 amsmath,amssymb,
 aps,
pra,
]{revtex4-1}

\usepackage{graphicx}
\usepackage{physics}

\newcommand{\rmi}{\mathrm{i}}
\newcommand{\rme}{\mathrm{e}}
\newcommand{\quantop}[1]{\mathcal{#1}}
\newcommand{\vect}[1]{\mathbf{#1}}

\newcommand{\boze}{\boldsymbol{\zeta}}

\begin{document}

\title{CV QKD with discretized modulations in the strong noise regime}

\author{Mikhail Erementchouk}%
 \email{merement@gmail.com}
 \author{Pinaki Mazumder}%
 \email{pinakimazum@gmail.com}
\affiliation{%
 Department of Electrical Engineering and Computer Science,\ 
 University of Michigan, Ann Arbor, MI 48109 USA 
}%

\begin{abstract}
We consider a general family of quantum key distribution (QKD) protocols utilizing displaced thermal states with discretized modulations. Separating the effects of the Gaussian channel and the non-Gaussian distribution, we have studied the dependence of the secret key generation rate on the magnitude of modulations (the strength of the modulated signal). We show that in the limit of strong signal, QKD is impossible: from the perspective of an efficient eavesdropper, the ensemble of transmitted states is effectively classical. This constitutes a quantum correction to performance of finite-length QKD protocols. We demonstrate that two regimes must be distinguished: weak and strong thermal noise. In the case of strong noise, the security boundary is mostly determined by the weak-signal limit. When the noise is weak, however, QKD may become possible only when the signal strength exceeds some critical value.
\end{abstract}

\maketitle

\section{Introduction}
\label{sec:intro}

The inherent asymmetry of three-way quantum communications is one of the drastic differences between classical and quantum communications. A quantum state sent by one party cannot be freely shared between remaining two. This circumstance is formalized by the famous no-cloning theorem: an unknown quantum state cannot be cloned \cite{scarani_quantum_2005,cerf_optical_2006}. Indeed, if such cloner existed, it would have to commute with all operators acting on the cloned state and, hence, its action would be independent of the cloned state. This demonstrates that the no-cloning property has similar fundamental roots as the Heisenberg uncertainty relation. Consequently, gaining information about an unknown state with necessity perturbs the state, as in  the noise-disturbance uncertainty relation \cite{ozawa_universally_2003}. Thus, roughly, sharing an unknown quantum state between two parties is a "zero-sum game": one party can gain information about the state only at the expense of another party.

This principal feature of quantum communications constitutes a foundation for the quantum key distribution (QKD) \cite{assche_quantum_2006,gisin_quantum_2002,weedbrook_gaussian_2012,diamanti_distributing_2015,laudenbach_continuous-variable_2018} aiming at producing by two parties probabilistically non-interceptable shared keys over authenticated channels. As hinted by the proof of the no-cloning theorem above, in order to avoid direct cloning, states that can be associated with non-commuting operators must be employed. Therefore, essentially, QKD protocols are based on sending non-orthogonal non-coinciding states and subsequent recovering of a shared key from apparently a random preparation and observation data. Since, at this stage, the data held by communicating parties is classical, it falls under classical Shannon's information framework and, hence, the shared key can be recovered using an adaptation of an error-correcting algorithm. 

Initially, QKD was developed for discrete variables, such as electron spin or photon polarization, but later the class of physical systems enabling QKD was extended by incorporating continuous variables (CV-QKD), for instance, quadratures of the electromagnetic field. Moreover, it was shown in Ref.\cite{weedbrook_continuous-variable_2012} that displaced thermal states can be used for generating the secret key thus dissociating QKD from the sole nature of utilized states. Since displaced states can be regarded as a result of quasi-classical driving of a cavity in thermal equilibrium, this significantly relaxes requirements with respect to state sources.

Bringing QKD to the realm of conventional sources boosted the development of practical QKD infrastructures, which potentially may significantly impact the field of secure communications. The main success in realizing CV-QKD is achieved in the optical and near-infrared spectral domains owing to the ready availability of highly coherent sources of the electromagnetic field and the low magnitude of thermal noise at room temperature \cite{laudenbach_continuous-variable_2018}.

The propagation of the QKD technologies further down the electromagnetic spectrum meets several obstacles. The main challenge appears to stem from thermal noise. With decreasing the base frequency, $\omega $, the noise magnitude grows fast, $\sim \exp(\omega_T/\omega) $ with $\omega_T = k_B T/\hbar $, once $\omega < \omega_T $. Here, $\hbar $ is the Planck constant, $k_B $ is the Boltzmann constant, and $T $ is the channel temperature. In Refs.~\onlinecite{weedbrook_quantum_2010,weedbrook_two-way_2014}, however, it was shown that strong thermal noise is not prohibiting on itself and rather determines the family of protocols (in this case, it is the direct reconciliation since it is more resistant with respect to noise).

Thus, further studies of QKD in the far-infrared and below spectral regions are warranted, motivated, on the one hand, by fundamental questions of the quantum/classical interface and the physical origin of information \cite{brillouin_science_1962}, and, on the other hand, by the demand to have matching technologies for emerging small-size high-bandwidth wireless networks.

In the present paper, we address a question that naturally arises in the context of low-frequency implementations of QKD. Main results with regard to the frequency dependence of QKD were obtained within the framework of Gaussian states, that is when the Wigner function of quantum states is a Gaussian function of field quadratures. Overall, this assumption is not too restrictive since the Gaussian property is preserved in dynamics governed by Hamiltonians quadratic in the field creation and annihilation operators. Such dynamics envelops a wide range of physical situations including linear and squeezing systems. However, in order for a train of transmitted state to submit to the formalism of Gaussian states, the variations of the transmitted states must follow the Gaussian distribution, in which case they essentially mimic thermal noise. In practical implementations, however, various deviations from the Gaussian distribution are unavoidable, which questions the applicability of the results obtained within the framework of Gaussian states. 

We consider the situation when the actual distribution of displacements of displaced thermal states is discretized,\footnote{Within the fields of conventional communications and signal processing, such signals are called quantized \cite{gallager_principles_2008} but for obvious reasons we will use less confusing terminology and call them discretized.} which clearly demonstrates deviations from the Gaussian framework. We revisit the standard theory of CV QKD for Gaussian states in order to distinguish effects inherent to Gaussian channels and those caused by the specific form of the distribution of the displacements. To this end, we have to abandon the convenient formalism of covariance matrices and to keep explicit operator form of relevant density matrices.

The strongest manifestations of the departure of discretized distributions from Gaussian is a non-monotonous dependence of the key generation rate on the intensity of the transmitted state. Moreover, the rate vanishes in the limit of strong excitations making QKD impossible. Physically, such reduction of the key generation rate can be understood as follows. Different states obtained by sufficiently strong displacement of thermal states are essentially orthogonal to each other and, thus, can be associated with (practically) commuting operators in the proof of the no-cloning theorem above. As a result, large values of the quantization parameter destroy the no-cloning character of the transmitted quantum states stripping the QKD off its fundamental background. It suggests that, in the QKD context, the transition to the classical regime emerges as an \emph{ensemble property} rather than that of individual states.

\section{CV QKD network with discretized modulations}

QKD protocols and networks are reviewed in a number of publications \cite{assche_quantum_2006,gisin_quantum_2002,weedbrook_gaussian_2012,diamanti_distributing_2015,laudenbach_continuous-variable_2018}. Therefore, here, we will limit ourselves to  setting up the problem of networks with discretized modulations and defining main notations without going into detailed discussion.

\subsection{The key generation rate}

In one-way QKD networks, the key is recovered from two strings of data held by the sender, $A $, and the receiver, $B $. On the $A $ side, the string $\Sigma_A  = \{\zeta_1, \ldots \}$ comprises the values of the control parameters, while, on the $B $ side, $\Sigma_B = \{\kappa_1, \ldots \}$ is populated by the results of observations. Assuming that there are no quantum correlations \emph{within} $\Sigma_A $ and $\Sigma_B $, these strings can be regarded as classical obtained as a result of a communication with abundant information over a noisy channel. According to Shannon's theory, the length of a perfectly correlated substring recoverable from $\Sigma_A $ and $\Sigma_B $ in the asymptotic limit is proportional to mutual information
\begin{equation}\label{eq:Iab}
 I(A:B) = \int d\zeta d \kappa \, \Pi(\zeta, \kappa) \ln \left[ \frac{\Pi(\zeta, \kappa)}{\Pi_0(\zeta) \Pi(\kappa)} \right],
\end{equation}
where $\Pi(\zeta, \kappa) $ is the joint distribution function of the controlling parameters and the results of observations, and $\Pi_0(\zeta) $ and $\Pi(\kappa) $ are the respective marginal distributions. The base of logarithm in Eq.~\eqref{eq:Iab} determines units for measuring information. We adopt natural units (nat), which slightly simplifies derived formulas. 

In one-way protocols, the distribution of outcomes of receiver's measurements deterministically depends on  transmitted state, so that the joint distribution has the form
\begin{equation}\label{eq:pixy}
 \Pi(\zeta, \kappa) = \Pi_{\quantop{K}}(\kappa | \zeta) \Pi_0(\zeta),
\end{equation}
where $\Pi_{\quantop{K}}(\kappa | \zeta) $ is the conditional probability of obtaining $\kappa $ while observing $\quantop{K} $ for a system in a state obtained with the controlling parameters set to $\zeta $. In physical terms, the conditional probability can be presented as $\Pi_{\quantop{K}}(\kappa | \zeta) = \Tr[\mathcal{E}_{\quantop{K}}(\kappa) \rho(\zeta) ] $, where $\rho(\zeta) $ is the density matrix of the full channel-environment state at the final stage of a QKD transaction starting from the state prepared with $\zeta $, and $\mathcal{E}_{\quantop{K}}(\kappa) $ is the respective spectral projector. Since, only the reduced density matrix at the receiving side is relevant, we have $\Pi_{\quantop{K}}(\kappa | \zeta) = \Tr[\mathcal{E}_{\quantop{K}}(\kappa) \rho_B(\zeta) ] $, where 
\begin{equation}\label{eq:B_side_dm}
 \rho_B(\zeta) = \Tr_E \left[ \rho(\zeta) \right],
\end{equation}
with traced out environmental degrees of freedom.

Using Eq.~\eqref{eq:pixy} in Eq.~\eqref{eq:Iab}, we obtain
\begin{equation}\label{eq:Iab_average}
I(A: B) = S \left( \overline{\Pi_{\quantop{K}}(\kappa |\zeta) } \right) - \overline{ S \left(\Pi_{\quantop{K}}(\kappa |\zeta)  \right)},
\end{equation}
where $S[f(\kappa)] = - \int \dd{\kappa} f(\kappa) \ln[f(\kappa)] $ is Shannon's entropy of distribution $f(x) $. The overline, as in Eq.~\eqref{eq:Iab_average}, denotes averaging with respect to the controlling parameter $ \overline{F(\zeta)} = \int \dd{\zeta} F(\zeta) \Pi_0(\zeta) $. For such averaging, we will also use the standard expectation symbol: $\mathbb{E} F(\zeta) = \overline{F(\zeta)} $. 

Applying an error correction kind of algorithm to $\Sigma_A $ and $\Sigma_B $, the communicating parties can ``recover the original message'' or, more formally, construct a common shared string $\Sigma_K $. In the absence of 
noise of uncontrolled origin (untrusted noise), $\Sigma_K $ would constitute a secret key. Thus, the rate of generation of secret key in this case is simply $R = I(A:B) $. In the presence of untrusted noise, however, the actual key must be constructed assuming that this noise is due to eavesdropping. In this case, the key rate must be adjusted to account for information intercepted by the eavesdropper, which yields
\begin{equation}\label{eq:key_rate}
 R = I(A:B) - \chi_E.
\end{equation}
Here, $\chi_E $ quantifies the amount of information accessible to the third party for a given magnitude of untrusted noise. Since $\Sigma_K $ is \emph{reconstructed} from $\Sigma_A $ and $\Sigma_B $ rather than transmitted, say, from $A $ to $B $, either $A $ or $B $ can be regarded as the holder of the ``original message'' and, respectively, either $A $ or $B $ can initiate error correction. These scenarios are called \emph{direct} and \emph{reverse reconciliation}, respectively \cite{grosshans_reverse_2002,grosshans_virtual_2003}. In the present paper, we limit ourselves to the case of direct reconciliation, as it demonstrates stronger resilience with respect to thermal noise. In this case, the maximum information is limited from above by the mutual quantum information between $A $ and $E $ (Holevo bound), $\chi_E = \chi(A:E) $ with
\begin{equation}\label{eq:holevo_AE}
 \chi(A:E) = H \left( \overline{\rho_E(\zeta)} \right) - \overline{ H \left( \rho_E(\zeta) \right)},
\end{equation}
where $H(\rho) = - \Tr[\rho \ln(\rho)] $ is the von Neumann entropy of the density matrix $\rho $ and $\rho_E(\zeta) = \Tr_B [\rho(\zeta)]$ is the density matrix of environment obtained by tracing out the receiver degree of freedom.

It must be noted that the fraction of recoverable message in a noisy string reaches Shannon's limit, $I(A:B) $, only asymptotically, when the length of the transmitted messages, $N$, is infinite, and the error correction algorithm is perfect. For finite $N$ and realistic algorithms, one needs to take into account that the recoverable message is shorter then prescribed by Shannon's limit. In the analysis of QKD protocols, this circumstance is accounted for by renormalizing the mutual information by the reconciliation efficiency $\lambda$, so that the actual secret key generation rate is given instead by $R = \lambda I(A:B) - \chi_E$. Usually, the reconciliation efficiency is regarded as determined by classical parameters and post-processing, see, e.g. \cite{jouguet_long-distance_2011,ruppert_long-distance_2014,jouguet_high-bit-rate_2014}. We, however, show below that there are corrections of essentially quantum origin that modify the key generation rate, so that the finite-$N$ effect  cannot be accounted by the reconciliation efficiency alone. Because of this circumstance, we will presume that main limitations arise due the discrete character of the displacement parameter and will take $\lambda = 1$.

\subsection{Transmitted states}

In the present paper, we limit ourselves to the single mode approximation, which assumes that only one mode contributes into QKD transactions.  First, we describe a general model of transmitted displaced single-mode states and establish general relations between these states and the mutual information that they can carry.

Displaced states are a particular case of Perelomov's coherent states \cite{perelomov_generalized_1986}. Let the sender's source cavity subjected to a semi-classical excitation be initially in thermal state 
\begin{equation}\label{eq:thermal_state}
\widetilde{\rho}(0; \widetilde{n}) = \frac{\rme^{-\beta a_0^\dagger a_0}}{1 + \widetilde{n}},
\end{equation}
where $\beta = \ln(1 + \widetilde{n}^{-1}) $, $\widetilde{n} $ is the average population of the cavity mode, and $a_0^\dagger $ and $a_0 $ are the cavity mode creating and annihilating operators, respectively. The dynamics of the driven cavity is described by $\quantop{H}_{int} = a_0^\dagger E + a_0 E^* $, where  $E $ is the complex amplitude of the external classical field. The evolution operator describing the action of the semi-classical excitation is Glauber's displacement operator $\quantop{D}_A $ and, thus, we assume that the states leaving the cavity have the form
\begin{equation}\label{eq:transmitted_general}
 \widetilde{\rho}_A(\widetilde{\zeta}; \widetilde{n}) = \mathcal{D}_A(\widetilde{\zeta}) \widetilde{\rho}_A(0; \widetilde{n}) \mathcal{D}_A^\dagger(\widetilde{\zeta}),
\end{equation}
with 
\begin{equation}\label{eq:displ_def}
 \mathcal{D}_A(\widetilde{\zeta}) = \exp \left( a_0^\dagger \widetilde{\zeta} - a_0 \widetilde{\zeta}^* \right).
\end{equation}
Here $\widetilde{\zeta} $ depends on the magnitude and duration of the classical driving field. Its relation with the displacement of transmitted states is described below in the model of discretized modulations.

The linear coupling between the channel mode and environment is described by the Hamiltonian $ \quantop{H}_e = f(t)\left( a_e^\dagger a_0 + a_0^\dagger a_e \right) $, where $a_e $ and $a_e^\dagger $ are the operators corresponding to  the external field. Let the initial state of the channel and the external field be $\widetilde{\rho}_c $ and $\rho_e $, respectively. Then, the result of such coupling is given by $\rho = \quantop{S} \widetilde{\rho}_c \otimes \rho_e \quantop{S}^\dagger $, where $\quantop{S} $ is the evolution operator describing the action of $\quantop{H}_e $. In order to describe the action of $\quantop{S} $, it is convenient to consider the external and the channel modes on the equal footing and to introduce vector notations $\vect{a}^\dagger \cdot \vect{v} \equiv v_0 a_0^\dagger + v_e a_e^\dagger $ with complex $v_0 $ and $v_e $. Then, the action of $\quantop{S} $ can be represented as
\begin{equation}\label{eq:linear_action}
 \quantop{S} f(\vect{a}^\dagger \cdot \vect{v} ) \quantop{S}^\dagger = f \left[ \vect{a}^\dagger \cdot (\widehat{S} \vect{v}) \right],
\end{equation}
where $\widehat{S} $ is the scattering matrix relating initial and final operators
\begin{equation}\label{eq:BS_mapping}
 \mqty( a_0(out) \\ a_e(out)) =
 	\widehat{S} \mqty( a_0(in) \\ a_e(in)),
\end{equation}
with
\begin{equation}\label{eq:BS_scattering}
 \widehat{S} = \mqty( t & r^* \\ -r & t^*).
\end{equation}
Thus, the linear coupling can be represented as mixing the external and channel mode on a beam-splitter characterized by complex reflection and transmission coefficients, $r $ and $t $, constrained by the unitarity condition $|t|^2 + |r|^2 = 1 $.

Measurements of the channel field after such interaction are described by the effective channel density matrix obtained by tracing the external degrees of freedom ${\rho}_c = \Tr_e \left[ \rho \right] $. If the channel is initially in the displaced thermal state $\widetilde{\rho}_c = \quantop{D}(\widetilde{\zeta}) \widetilde{\rho}_A(0; \widetilde{n}) \quantop{D}^\dagger(\widetilde{\zeta}) $, then ${\rho}_c $ is also a displaced thermal state
\begin{equation}\label{eq:map_displaced_state}
 {\rho}_c = \quantop{D}(\zeta) {\rho}_A(0) \quantop{D}^\dagger(\zeta),
\end{equation}
where $\zeta = \vect{e}_0^\dagger \cdot (\widehat{S} \boze)$ with $\vect{e}_0^\dagger = (1, 0) $ and $\boze = (\widetilde{\zeta}, 0)^T $, so that $\zeta = t \widetilde{\zeta} $, and
\begin{equation}\label{eq:modified_r_A}
 {\rho}_A(0) = \Tr_e \left[ \quantop{S} \widetilde{\rho}_A(0) \otimes \rho_e \quantop{S}^\dagger \right].
\end{equation}

Let the ambient electromagnetic field be in a thermal state characterized by the average population $n_a$. The beam splitter turns the incoming state into the channel state (see Stage I in Fig.~\ref{fig:overall})
\begin{equation}\label{eq:channel_state}
 \rho_A(\zeta; n) = \Tr_a \left[ \quantop{S}(r,t) \rho_A(\widetilde{\zeta}; n_0) \otimes \rho_{th}(n_a) \quantop{S}^\dagger(r,t) \right],
\end{equation}
where $\quantop{S}(r,t) $ is an operator describing the transformation induced by the beam splitter.

Using the $P $-representation for the density matrices, we can rewrite this equation as
\begin{equation}\label{eq:channel_state_expanded}
\begin{split}
\rho_A(\zeta; n) = & \frac{1}{\pi^2 n_0 n_a} \int d^2 z_0 d^2 z_a \rme^{-|z_0|^2/n_0 -|z_a|^2/n_a} \\
 & \times \Tr_a \left[ \quantop{S} \quantop{D}(\vect{v}) \ket{0} \bra{0} \quantop{D}^\dagger(\vect{v}) \quantop{S}^\dagger  \right],
\end{split}
\end{equation}
where $\quantop{D}(\vect{v}) = \exp(\vect{v} \cdot \vect{a}^\dagger - \vect{v}^* \cdot \vect{a}) $ and $\vect{v}\cdot \vect{a}^\dagger = v_0 a_0^\dagger + v_a a_a^\dagger $ with $v_0 = z_0 + \zeta $ and $v_a = z_a $. Taking into account that $\quantop{S} \quantop{D}(\vect{v}) \ket{0} = \quantop{D}(\vect{u}) \ket{0} $ with $\vect{u} = \widehat{S} \vect{v} $, we obtain
\begin{equation}\label{eq:channel_state_red}
\begin{split}
 \rho_A(\zeta; n) = \frac{1}{\pi^2 n_0 n_a} & \int d^2 z_0 d^2 z_a \rme^{-|z_0|^2/n_0 -|z_a|^2/n_a}\\
  & \times \quantop{D}(u_0) \ket{0} \bra{0} \quantop{D}^\dagger(u_0).
\end{split}
\end{equation}
By changing the integration variables, Eq.~\eqref{eq:channel_state_red} can be turned into the canonical form yielding
\begin{equation}\label{eq:zeta_n_eff}
\zeta = t \widetilde{\zeta}, \qquad n = |t|^2 n_0 + |r|^2 n_a.
\end{equation}
Thus, the modulation of the transmitted state for a given outcome of the source of displaced thermal states can be achieved by varying the complex transmission coefficient of the beam splitter. The modulation of post-source states is commonly used in experimental implementations of QKD.

If controls determining the value of $t $ admit a finite number of states, $t $ takes values at a finite number of points inside the unit circle on the complex plane. These points are mapped by multiplication by $\widetilde{\zeta} $ into the complex $\zeta $-plane resulting in discretized modulations. In the present paper, we consider the effect of the magnitude of $\widetilde{\zeta} $ or, more physically, of the strength of the quasi-classical excitation, on the key generation rate. To this end, we represent the modulation value as $s \zeta $, where $s $ is a scaling parameter.

Some results obtained below can be formulated for a general observable $\quantop{K} $ measured at the receiving end. Such generalization may be of interest in the context of low-frequency spectral domains, where a wide variety of methods to control the electromagnetic field is available. In the present paper, however, we will limit ourselves to the case when quadratures are measured. In this case, the conditional probability to obtain value $\kappa $ is given by
\begin{equation}\label{eq:quadrature_kernel}
 \Pi_{\quantop{K}}(\kappa | \zeta) = Q(\kappa | \zeta ) \equiv \frac{1}{\sqrt{2 \pi \sigma^2}} 
 	\exp \left\{ - \frac{1}{\sigma^2} \left[ \kappa - \expval{\kappa_\zeta} \right]^2 \right\},
\end{equation}
where $\sigma^2 = 2n + 1 $ and $\expval{\kappa_\zeta} = \sqrt{2} \Re \left( t \zeta \rme^{\rmi \theta} \right) $. The family of quadratures is parametrized by the phase parameter $\theta $ and the argument of the channel transmission coefficient. A variety of protocols  is based on the precise control over the quadrature phase provided by synchronizing the local oscillator in the homodyne detection of the quadrature. Here, we do not put any restrictions on the phase thus allowing for an unsynchronized local oscillator.

\subsection{Physical model of the information loss}

We model an efficient coupling with environment using the model of Gaussian collective attacks. These attacks are proven to be optimal for Gaussian protocols and are conjectured to be optimal in general \cite{navascues_optimality_2006,garcia-patron_unconditional_2006,pirandola_characterization_2008}. Within this model, eavesdropping masks itself as thermal noise, so that the initially the external coupled state purifies thermal state. More specifically, the external field is initially prepared in a two-mode squeezed vacuum (TMSV) state
\begin{equation}\label{eq:TMSV}
 \rho_E^{(0)} = \quantop{F}(\mu) \ket{0} \bra{0} \quantop{F}^\dagger(\mu),
\end{equation}
where $\quantop{F}(\mu) $ is the two-mode squeezing operator. Denoting the operators of the environment modes by $a_2 $ and $a_3 $, we have
\begin{equation}\label{eq:tms_operator}
 \quantop{F}(\mu) = \exp \left[ \mu \left( a_2^\dagger a_3^\dagger - a_2 a_3 \right) \right].
\end{equation}
Generally, the squeezing parameter can be complex. Its argument, however, can be absorbed into $a_{2,3} $ without changing final results. Therefore, Eq.~\eqref{eq:tms_operator} presumes that the squeezing parameter is a real number, which simplifies intermediate formulas.

\begin{figure}[tb]
	\centering
	\includegraphics[width=3.5in]{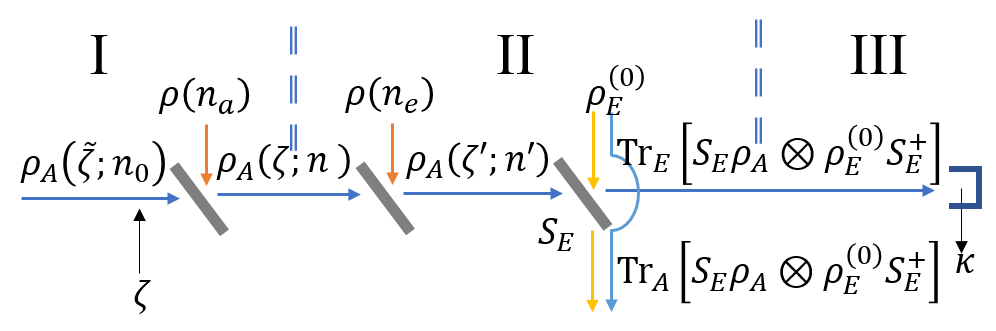}
	\caption{Propagation of quantum state in a QKD transaction. Stage I: preparation of quantum state. Stage II: the effect of environment and the model of information losses. Stage III: detection.}
	\label{fig:overall}
\end{figure}

One of the squeezed modes is mixed with the channel mode on a beam splitter, while the second mode is collected together with the mode transmitted through the beam splitter (see Stage II in Fig.~\ref{fig:overall}), which constitutes $\rho_E(\zeta) $ in Eq.~\eqref{eq:holevo_AE}. The strength of coupling of the channel mode with environment is quantified by the reflection coefficient of the beam splitter, $r_E $, which also can be assumed real without loss of generality.

\section{Untrusted noise and information leaked into environment}

One of the main objectives of a theory of QKD is to establish the amount of leaked information for a given (measured during the communication session) amount of untrusted noise. Based on this knowledge, the communicating parties decide whether the secret key can be extracted (if mutual information exceeds losses) or the results of the communication session must be abandoned.

\subsection{The emergence of untrusted noise}

An efficient coupling of channel modes with environment in a purified thermal states affects the channel mode in the same way as coupling with a thermal state. Indeed, while evaluating the partial trace over the eavesdropper's modes in Eq.~\eqref{eq:B_side_dm}, one needs to take into account that the displacement direction of the channel state is orthogonal to the plane of squeezing. Thus, tracing out the mode, which is not mixed with the state in the channel, yields
\begin{equation}\label{eq:trace_out_E_Fock_3}
 \rho_B(\zeta)  = \sum_{m_2} \matrixel{m_2}{\quantop{S}(\theta) \rho_{A}(\zeta; n) \otimes 
\widetilde{\rho}_{th} \quantop{S}^\dagger(\theta)	 }{m_2} ,
\end{equation}
where $\widetilde{\rho}_{th}$ is an effective thermal state 
\begin{equation}\label{eq:effect_th}
\widetilde{\rho}_{th} = \frac{1}{\cosh^2(\mu) } \sum_{m_3} \tanh^{2m_3}(\mu) \ket{m_3}\bra{m_3}.
\end{equation}
Thus, from the channel perspective, the efficient coupling is indistinguishable from coupling with a thermal state $\rho_{th}(\sinh^2(\mu)) $. If all environment states are of uncontrolled origin, then at the receiving side we have $\rho_B(\zeta) = \rho_{th}(t_E \zeta, n t_E^2 + n_E) $, where 
\begin{equation}\label{eq:ne_def}
{n}_E = r_E^2 \sinh^2(\mu)
\end{equation}
is the magnitude of untrusted noise and we have taken into account that the parameters describing the strength of the coupling with environment, $t_E $ and $r_E $, can be chosen real.

Importantly, this implies the reverse: any untrusted noise must be regarded as stemming from the information loss through the efficient coupling.

\subsection{Information loss}

When $A$ announces its data, the amount of leaked information is limited by the Holevo bound $\chi(A:E) $. Since we do not assume the Gaussian form of $\Pi_0(\zeta) $, it is convenient to rewrite the environment density matrix in the form distinguishing non-Gaussian modulations and propagation in the Gaussian channel
\begin{equation}\label{eq:rho_E_red}
 \rho_E(\zeta) = \quantop{D}_2(-r \zeta) \widetilde{\rho}_E \quantop{D}_2^\dagger (-r \zeta),
\end{equation}
where
\begin{equation}\label{eq:rho_E_TMSV}
 \widetilde{\rho}_E = \Tr_A \left[ \quantop{S}(\theta) \rho_{th}(n) \otimes \rho_{\quantop{F}}(\mu) \quantop{S}^\dagger(\theta) \right]
\end{equation}
is the density matrix of a TMSV mixed with a thermal state. This density matrix is independent of the modulation parameter and, due to invariance of von Neumann entropy with respect to unitary transformations of the density matrix, we immediately obtain
\begin{equation}\label{eq:average_S}
  \overline{ H \left( \rho_E(\zeta) \right)} = H \left( \widetilde{\rho}_E \right).
\end{equation}

While $\widetilde{\rho}_E $ is a Gaussian state and, therefore, is completely characterized by its covariance matrix, in order to find $\overline{\rho_E(\zeta)} $, it is convenient to have an explicit form of $\widetilde{\rho}_E $ in an operator form. It can be recovered from the covariance matrix. We find it constructive, however, to perform the calculation using the representation in terms of creation and annihilation operators and to demonstrate the emergence of the phase space representation. It can be done, for example, as follows. Using the $P$-representation for $\rho_{th} $ in Eq.~\eqref{eq:rho_E_TMSV}, it can be rewritten as $\widetilde{\rho}_E = \quantop{F}(\mu_t) \widehat{\rho}_E \quantop{F}^\dagger(\mu_t)$, where $\mu_t $ is defined by $\tau_t \equiv \tanh (\mu_t) = t \tanh(\mu) $ and
\begin{equation}\label{eq:stripped_E}
\begin{split}
 \widehat{\rho}_E = & \frac{1}{\pi \bar{n}} \int \dd{\alpha} \rme^{- |\alpha|^2/\bar{n}} 
 					\quantop{D}_2(\alpha_c) \quantop{D}_3(\alpha_s) \\
 & \times \ket{0}\bra{0} \otimes \rho^{(3)}_{th}(\bar{n}_E) 
 			\quantop{D}_3^\dagger(\alpha_s) \quantop{D}_2^\dagger(\alpha_c),
\end{split}
\end{equation}
with $\alpha_c = r \alpha \cosh(\mu_t) $, and $\alpha_s = - r \alpha^* \sinh(\mu_t) $. In this expression, $\rho^{(3)}_{th}({n}_E) = Z_E^{-1} \exp(-\beta_E a_3^\dagger a_3) $ is a thermal state characterized by the same average number of particles $ {n}_E = r^2 \sinh^2(\mu) $ as the magnitude of untrusted noise. Using again the $P $-representation turns Eq.~\eqref{eq:stripped_E} into
\begin{equation}\label{eq:stripped_E_2}
\begin{split}
 \widehat{\rho}_E = & \frac{1}{\pi^2 {n}_{r} {n}_E} \int \dd{\vect{z}} 
 					\rme^{-|z_2|^2/{n}_{r} -|z_3 + \tau_t z_2^*|^2/{n}_E} \\
 				&	\quantop{D}(\mathbf{z}) \ket{0} \bra{0} \quantop{D}^\dagger(\mathbf{z}),
\end{split}
\end{equation}
where we have introduced ${n}_{r} = r^2 {n} \cosh^2(\mu_t) = r^2 n/(1 - \tau_t^2) $, $\dd{\vect{z}} = \dd{z_2} \dd{z_3}$, and
\begin{equation}\label{eq:displ_23}
 \quantop{D}(\vect{z}) = \exp(z_2 a_2^\dagger + z_3 a_3^\dagger - \mathrm{h.c.}).
\end{equation}
A connection with the phase space formalism is then established through Williamson's theorem \cite{williamson_algebraic_1936} that guarantees that any Gaussian state can be presented as a transformation of a direct product of thermal states. In terms of representation of the density matrix given by Eq.~\eqref{eq:stripped_E_2}, this means that the form $-|z_2|^2/{n}_{r} -|z_3 + \tau_t z_2^*|^2/{n}_E $ can be diagonalized by proper transformations. To this end, it is convenient to rewrite the argument in Eq.~\eqref{eq:displ_23} as
\begin{equation}\label{eq:pre_sympl}
 \vect{z} \cdot \vect{a}^\dagger - \vect{z}^* \cdot \vect{a} = 
  \mqty(\vect{z} & \vect{z}^*) \widehat{J} \mqty(\vect{a} \\ \vect{a}^\dagger),
\end{equation}
where $ \widehat{J} =  \mqty(0 & \widehat{1} \\ -\widehat{1} & 0)$, with $\widehat{1} $ being the $2\times 2 $ identity matrix, is a symplectic form consistent with the commutation relations: $\quantop{C} - \quantop{C}^T = \widehat{J}  $, where $\quantop{C} = \mqty(\vect{a} \\ \vect{a}^\dagger) \otimes \mqty(\vect{a} \\ \vect{a}^\dagger) $. It can be seen that transformations of creation and annihilation operators preserving the commutation relations induce ``symplectic orthogonal'' transformation of $z_{1,2} $. Indeed, transformation of operators $\vect{a} \to \vect{b} $ according to
\begin{equation}\label{eq:op_trans}
 \widehat{R} \mqty(\vect{a} \\ \vect{a}^\dagger) = \mqty(\vect{b} \\ \vect{b}^\dagger)
\end{equation}
induces transformation $\vect{z} \to \vect{w} $
\begin{equation}\label{eq:z_trans}
 - \mqty(\vect{z} & \vect{z}^*) \widehat{J} \widehat{R} \widehat{J} = \mqty(\vect{w} & \vect{w}^*).
\end{equation}
For example, two-mode squeezing described by operator $\quantop{F}(\gamma) $ yields
\begin{equation}\label{eq:w_squeezing}
 \begin{split}
 w_2 = z_2 \cosh(\gamma) - z_3^* \sinh(\gamma), \\
 w_3 = z_3 \cosh(\gamma) - z_2^* \sinh(\gamma).
 \end{split}
\end{equation}
It turns out, two-mode squeezing is the only transformation needed for diagonalization of the form in the exponential term in Eq.~\eqref{eq:stripped_E_2}, so that
\begin{equation}\label{eq:stripped_E_3}
 \widehat{\rho}_E = \quantop{F}(\gamma) \rho^{(2)}_{th}({n}_2) \otimes \rho^{(3)}_{th}({n}_3) \quantop{F}^\dagger(\gamma)
\end{equation}
where 
\begin{equation}\label{eq:tan_angl}
\tanh(2 \gamma) = \frac{2 \tau_t}{Y + 2}
\end{equation}
with $Y = X + \tau_t^2 -1 $ , $X = {n}_r/{n}_E $, and
\begin{equation}\label{eq:eff_ns}
 {n}_{2,3} = \frac{2 {n}_E}{\sqrt{Y^2 + 4 X} \pm Y}.
\end{equation}

Collecting these results, we obtain the averaged density matrix of environment (up to a $\zeta $-independent unitary transformation) $\overline{\rho_E} = \int \dd{\zeta} \Pi_0(\zeta) \rho_E(\zeta) $, where
\begin{equation}\label{eq:effective_environment_dm}
\rho_E(\zeta) = \quantop{D}(\vect{z}(\zeta)) 
  			\rho^{(2)}_{th}(n_2) \otimes \rho^{(3)}_{th}(n_3) \quantop{D}^\dagger(\vect{z}(\zeta)) ,
\end{equation}
with $z_2(\zeta) = - s \zeta r_E \cosh(\mu + \gamma) $ and $z_3(\zeta) = s \zeta^* r_E \sinh(\mu + \gamma) $.

\section{Key generation rate}

\subsection{Strong modulation limit}

\begin{figure}[tb]
	\centering
	\includegraphics[width=2.8in]{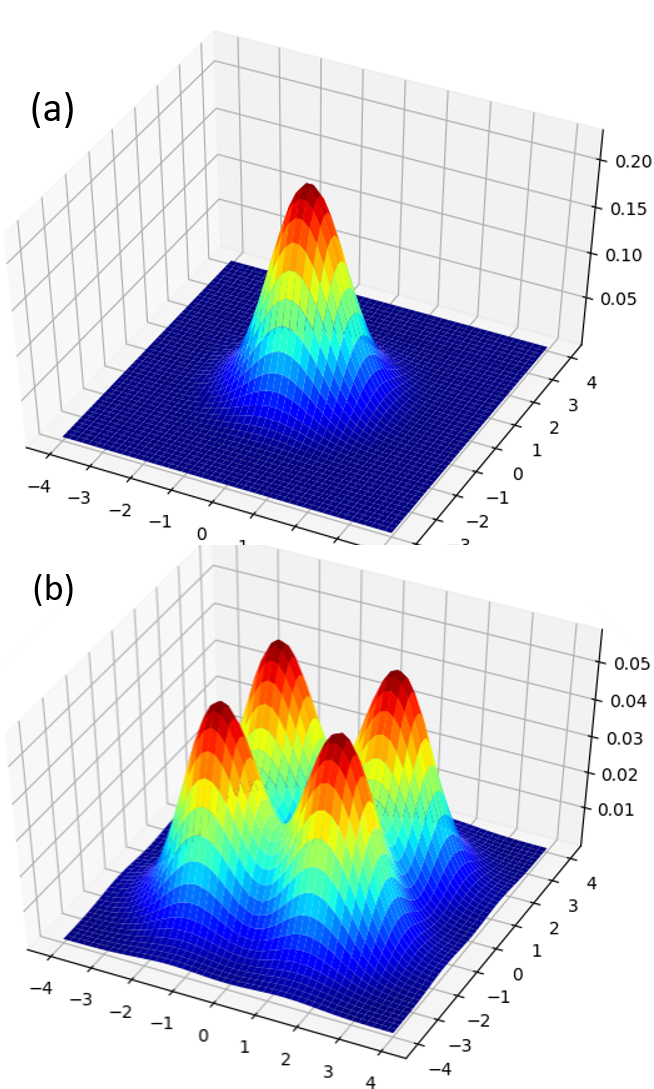}
	\caption{An example of the Wigner distribution of the environment density matrix traced over one of the degrees of freedom, $\rho^{(red)}_E = \Tr_2[\widetilde{\rho}_E] $ in the case of (a) weak and (b) strong modulations. The distribution of $\zeta $'s is assumed to be uniform over points $\pm 1 \pm \rmi $ of the complex plane.}
	\label{fig:wigners}
\end{figure}

The limit of strong modulations, when the magnitude of $\zeta $ exceeds characteristic scales describing channel and environment modes, is the simplest since, in this limit, quantum correlations in environment between individual transactions become negligible. Figure~\ref{fig:wigners}(a) shows partial Wigner distribution of the environment density matrix in the case when the modulation parameter is strong $s \gg 1 $, so that the fluctuations of the displacement parameter exceed the width of Gaussian states $s \Delta_\zeta \gg \max(n_{2}, n_{3})  $, where $n_{2, 3} $ are given by Eq.~\eqref{eq:eff_ns}, and $\Delta_\zeta $ is the magnitude of a ``typical'' separation between points in the $\zeta $-plane. The multimodal character of the Wigner distribution in Fig.~\ref{fig:wigners}(b) is the principal feature of the environment density matrix when the magnitude of discretized modulations becomes too strong.

A formal manifestation of this observation is vanishing commutators of individual terms with $\zeta \ne \zeta' $ in $\overline{\rho_E} $
\begin{equation}\label{eq:comm_est}
 \left[ \rho_E(\zeta), \rho_E(\zeta') \right] \propto \rme^{ - (\vect{z}(\zeta) - \vect{z}(\zeta'))^*
 		\cdot \widehat{n}^{-1} \cdot (\vect{z}(\zeta) - \vect{z}(\zeta'))},
\end{equation}
where $ \widehat{n} = \mathrm{diag}(n_2, n_3)$. Based on that, the perturbation theory can be used for an analysis of the spectrum of $\overline{\rho_E} $ with the characteristic decay of small terms $\propto \rme^{- s^2/s_0^2} $ with $s_0^{-2} \propto \Delta_\zeta r_E^2 \left( n_2^{-1} \cosh^2(\mu + \gamma)  + n_3^{-1} \sinh^2(\mu + \gamma)\right)  $ as $s \to \infty $. The precise form of $s_0$ depends on the mutual arrangement of eigenvalues of individual terms in $\overline{\rho_E} $. For example, when $\Pi_0(\zeta) = 1/M $, where $M $ is the total number of values of modulations, all eigenvalues of $\Pi_0(\zeta) \rho_E(\zeta) $ are $M$-fold degenerate and $s_0^2 $ may acquire a factor depending on details of distribution of $\zeta $'s in the complex plane. As will be apparent from the following, however, the exact asymptotic form of the Holevo bound may be of rather minor importance. Therefore, for the purpose of the present paper it suffices to limit ourselves to the zeroth order of the perturbation theory, when the overlap between the eigenstates of $\rho_E(\zeta) $ and $\rho_E(\zeta') $ for $\zeta \ne \zeta' $ is completely neglected.

In this case, the commutator above vanishes and the environment density matrix reduces to the direct sum of individual $\rho_E(\zeta) $. Taking into account that $ H(\bigoplus_n \rho_n) = \sum_n H(\rho_n) $ for any set of commuting operators $\rho_n $. and $H(a \rho) = - a \ln(a) + a H(\rho) $ for a real number $a $ and normalized $\rho $, we obtain
\begin{equation}\label{eq:h_of_av}
 H \left( \overline{\rho_E} \right) = S \left( \Pi_0(\zeta) \right) + \overline{H(\rho_E(\zeta))}.
\end{equation}
Thus, in the limit of strong modulations, the Holevo bound saturates at the entropy of the distribution of the modulation parameter. Since this entropy limits the amount of transmitted information, we conclude that in the limit of strong modulations the rate of generation of the secure key is vanishing and the QKD is impossible.

At the same time, it should be noted that the actual limit reached by $I(A:B) $ may be strictly smaller than $ S \left( \Pi_0(\zeta) \right) $ owing to the details how the modulation parameter enters the propagator $Q(\kappa | \zeta) $. For example, as shown in Eq.~\eqref{eq:quadrature_kernel}, the homodyne detection of quadratures depends on the value of modulation through $\expval{\kappa_\zeta} = \sqrt{2} \Re(t \zeta \rme^{\rmi \theta}) $. Then, in the limit of strong modulations, the mutual information asymptotically tends to the entropy of distribution of this parameter, $I(A:B) = S(\Pi_0 (\expval{\kappa_\zeta})) $. Up to scaling, the distribution of $\expval{\kappa_\zeta} $ has the same form as that of the projection of the distribution of $\zeta $ onto the line passing through the origin of the complex plane at the angle determined by the angular parameter of the quadrature and the phase of the effective transmission coefficient $t $. If the distribution of $\zeta $ has a cluster form, after such projection the clusters may overlap yielding a distribution with smaller entropy.

From the QKD perspective, the consequence of mismatched asymptotics of $I(A:B) $ and $ \chi(A:E)$ is that in the limit of strong modulations there is a sharp security boundary: there is a maximal magnitude, beyond which QKD is impossible.

It should be noted that, in the consideration above, the quadrature phase parameter $\theta$ is not presumed to be controlled by communicating parties. Such control can be achieved by synchronizing the local oscillator in the homogeneous detection of quadratures. On the one hand, this provides means to ensure the certain orientation of distribution $\Pi_0(\zeta)$ in the complex plane thus minimizing the loss of information due to its projection on the real axis. On the other hand, due to the effect of the phase acquired during propagation, accounted for by the argument of the effective transmission coefficient $t$, such synchronization is a non-trivial task and poses a challenge for practical implementations of CV QKD. Therefore, it is worth noting that the analysis above confirms that such synchronization, while beneficial, is not strictly required \cite{soh_self-referenced_2015}. Random variations of $\theta$ can be taken into consideration while optimizing particular implementations and accounted for in the estimate of the mutual information between the communicating parties.

We take into account this circumstance by limiting ourselves in the following numerical evaluations to distributions $t \zeta \rme^{\rmi \theta} $ confined to the real axis. This doesn't impact significantly the generality, while simplifies the discussion.

Because the physical origin of vanishing key generation rate is the effective emergence of the classical ensemble of states due to the weak overlap of individual density matrices in the limit of strong variation, it affects all protocols based on displaced coherent states, including those with Gaussian distribution of the displacing parameter. Because of the finite length of the sequence of transmitted quantum states, the signal of sufficiently strong amplitude will ``separate'' individual states leading to collapsing key generation rate. It must be noted that this kind of finite-length effect cannot be accounted by reconciliation efficiency, which quantifies the error correction algorithm and renormalizes the mutual information. Moreover, high efficiency (yielding $\lambda > 0.95$) is reached in the limit of high signal-to-noise ratio \cite{jouguet_long-distance_2011,ruppert_long-distance_2014,jouguet_high-bit-rate_2014}, thus making the estimate of the protocol performance vulnerable with respect to the effect of emergence of classical ensembles when the length of the sequence of transmitted quantum states is relatively small.

\subsection{Weak modulation limit}

In the opposite limit of weak modulations (small $s$), both the mutual information and the Holevo bound vanish in a threshold-less manner and their Taylor expansions start with terms quadratic in $s$. Thus, in this limit, 
\begin{equation}\label{eq:r_weak}
 R = s^2 C,
\end{equation}
where 
\begin{equation}\label{eq:strong condition}
 C = \frac{d^2}{ds^2} \left[ I(A: B) - \chi(A:E) \right].
\end{equation}
The key can be generated, if $C > 0$.

It follows straightforwardly from Eq.~\eqref{eq:quadrature_kernel} that
\begin{equation}\label{eq:I_rate}
  \frac{d^2}{ds^2} I(A: B) = \frac{2}{\sigma^2} \mathbb{E} \left[ \expval{\kappa(\zeta)} - \overline{\expval{\kappa(\zeta)}} \right]^2.
\end{equation}
It should be noted, that, in this limit, the non-ideal reconciliation efficiency leads to simple renormalization $\expval{\kappa(\zeta)} \to \sqrt{\lambda} \expval{\kappa(\zeta)} $.

The Holevo bound is determined by the eigenvalues of $\rho_E$. When $s = 0$, they are given by the product of eigenvalues of $\rho_{th}^{(2, 3)} $ in Eq.~\eqref{eq:effective_environment_dm}. Since $\rho_{th}^{(2, 3)} $ are diagonal in the product of Fock bases, it is convenient to introduce a ``vector'' notation for the basis states $\ket{\vect{l}} \equiv \ket{l_2, l_3} $, so that $\rho_E^{(\vect{l})}(0) $, the eigenvalues at $s = 0$, can be expressed in terms of the average number of thermal photons $n_{2,3} $ as 
\begin{equation}\label{eq:s0_eigenvalues_env}
\rho_E^{(\vect{l})}(0)  = 
			\frac{\rme^{-\beta_2 l_2 - \beta_3 l_3} }{\left( 1 +  {n_2} \right) 
													\left( 1 +  {n_3} \right)},
\end{equation}
where $\beta_{2,3} = \ln(1 + 1/n_{2,3}) $.

Since we are interested only in the variation of the eigenvalues, we can use the similar approach as for the Feynman-Hellmann theorem. The first order is given by $\left. \partial \rho_E^{(\vect{l})} / \partial s \right |_{s = 0} = \matrixel{\vect{l}}{\partial \rho_E(0) / \partial s}{\vect{l}} $, while in the second order we have
\begin{equation}\label{eq:pert_rho}
\begin{split}
\left. \frac{\partial^2 \rho_E^{(\vect{l})}}{\partial s^2} \right |_{s = 0} & = \matrixel{\vect{l}}{\partial^2 \rho_E(0) / \partial s^2}{\vect{l}}
	+ \\
	& +  2 \sum_{\vect{m} \ne \vect{l}} 
		\frac{\matrixel{\vect{l}}{\partial \rho_E(0) / \partial s}{\vect{m}} 
		\matrixel{\vect{m}}{\partial \rho_E(0) / \partial s}{\vect{l}}}{\rho_{2,3}^{(\vect{l})}(0) - \rho_{2,3}^{(\vect{m})}(0)}
		,
\end{split}
\end{equation}
where $\vect{l} = (l_2, l_3) $ and $\vect{m}  = (m_2, m_3) $.

Introducing $ s \quantop{V}(\zeta) = \vect{z}(\zeta)\cdot \vect{a}^\dagger - \vect{z}^* (\zeta) \cdot \vect{a} $, these expressions can be rewritten in a more explicit form
\begin{equation}\label{eq:alt_pert_rho}
\begin{split}
 \left. \frac{\partial}{\partial s} \rho_E^{(\vect{l})} \right |_{s=0} = &
 		\matrixel{\vect{l}}{\left[ \overline{\quantop{V}(\zeta)}, \rho_E(0) \right] }{\vect{l}} ,\\
 \left. \frac{\partial^2}{\partial s^2} \rho_E^{(\vect{l})} \right |_{s=0} = &
 	\matrixel{\vect{l}}{\left( \overline{\quantop{V}^2(\zeta)} \rho_E(0) + 
 				\rho_E(0) \overline{\quantop{V}^2(\zeta)}\right) }{\vect{l}}- \\
	& -  2 \matrixel{\vect{l}}{\overline{\quantop{V}(\zeta) \rho_E(0) \quantop{V}(\zeta)} }{\vect{l}}.
\end{split}
\end{equation}

Because of invariance of the von Neumann entropy with respect to unitary transformations of the density matrix, we can set $\overline{\zeta} = 0 $ without any loss of generality, which yields
\begin{equation}\label{eq:HB_derivative}
\left. \frac{\partial^2 }{\partial s^2 } \chi(A:E) \right|_{s=0} = \sum_{\vect{l}} \frac{\partial^2 }{\partial s^2 } \rho_E^{(\vect{l})}(0)  \ln \left[ \rho_E^{(\vect{l})}(0) \right].
\end{equation}
Using Eq.~\eqref{eq:alt_pert_rho} in this expression, we obtain
\begin{equation}\label{eq:chi_curvature}
\begin{split}
\left. \frac{\partial^2 }{\partial s^2 } \chi(A:E) \right|_{s=0} = 
		2 \overline{|\zeta|^2} r_E^2 & \left[ \beta_2 \cosh^2(\mu + \gamma) \right. \\
						& \left.	+ \beta_3 \sinh^2(\mu + \gamma) \right].
\end{split}
\end{equation}
Together with Eq.~\eqref{eq:I_rate}, this expression gives an explicit condition whether the QKD is possible in the limit of weak modulations. 

We conclude consideration of the limiting cases by noticing that they imply that the key generation rate is a non-monotonous function of the signal strength. Thus, an implementation of a QKD protocol based on discretized modulations must include solution of the respective optimization problem taking into account the characteristics of the communication channel and the magnitude of untrusted noise.

\section{Weak and strong noise regimes}

\begin{figure}[tb]
	\centering
	\includegraphics[width=3.3in]{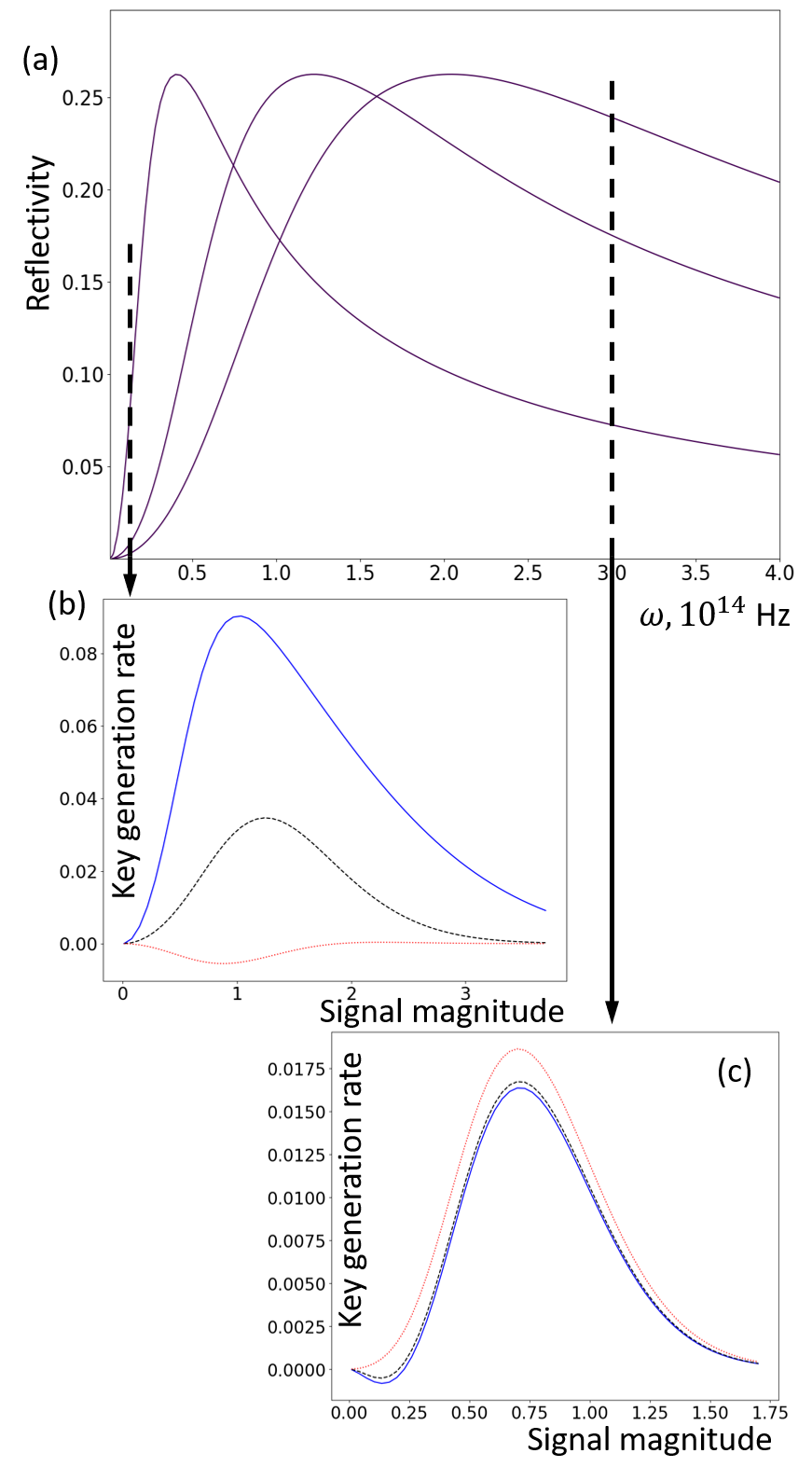}
	\caption{(a) The security boundary $C(r_E, \omega) = 0$ (see Eq.~\eqref{eq:strong condition}) in up to the mid-infrared region (the shortest wavelength is $4.7$ $\mu$m) for fixed effective temperature of environment: (1) $T = 100$ K, (2) $T= 300$ K, (3) $T = 500$ K. The regions above and below the curve correspond to insecure and secure regimes, respectively. (b, c) Numerical evaluation of the signal dependence of the key generation rate at (a) $\omega = 2\cdot 10^{13} $ Hz and $r_E^2 = 0.01 $, and (b) $\omega = 3\cdot 10^{14} $ Hz and $r_E^2 = 0.22 $. }
	\label{fig:sec_strong}
\end{figure}

In order to investigate the dependence of the security boundary given by $C = 0$ on parameters of environment and coupling with it, we assume that the effective temperature of environment is fixed. In Fig.~\ref{fig:sec_strong}(a), we plot $C(r_E, \mu, \omega) = 0 $, the phase diagram separating secure and insecure regimes, with imposed constrain $ r_E^2 \sinh^2(\mu) = \mathrm{const} $, as a function of coupling with environment and the carrier frequency (energy) of the quantum states. It demonstrates that at low frequencies, the security boundary obtained in the weak modulation limit correctly distinguishes secure and insecure regimes even when the modulation is not necessarily weak. Presented in Fig.~\ref{fig:sec_strong}(b) signal dependencies of the key generation rate show that the sign of $R$ does not change with the magnitude of displacement. 

\begin{figure}[tb]
	\centering
	\includegraphics[width=3.3in]{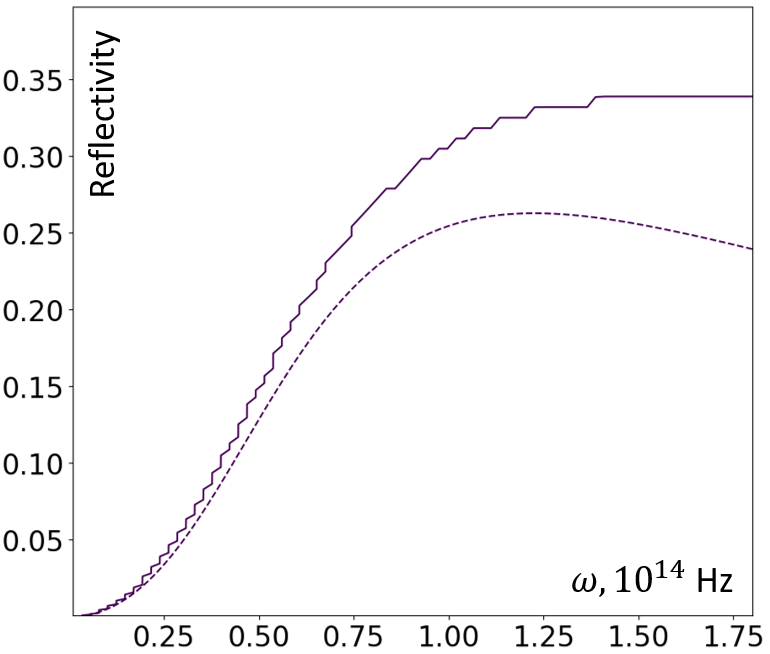}
	\caption{The security boundary on the $(r_E, \omega) $-plane for $T = 300$ K. The dashed line shows the security boundary based on the weak signal approximation and the solid line presents the security boundary obtained by a numerical simulation of QKD transactions with discretized modulations (dotted line).}
	\label{fig:sec_regimes}
\end{figure}

The security boundary defined as $C = 0$ predicts that with increasing frequency the maximal coupling with environment admitting generation of the key eventually starts to decrease signifying that the condition $C = 0$ is no longer applicable when the number of thermal photons at the energy of transmitted states becomes \emph{smaller} than one. This observation is confirmed by comparing the security boundary found as $C = 0$ with numerically obtained security boundary presented in Fig.~\ref{fig:sec_regimes}.

It should be noted that the condition $C = 0$ correctly predicts the security of protocols utilizing weak states even in this case. To illustrate this circumstance, we show in Fig.~\ref{fig:sec_strong}(c) the signal dependence of the key generation rate in the the case when thermal noise is small. It shows that for systems that are in different regions according to the weak signal and a precise condition, there is a critical magnitude of the signal, below which QKD is impossible. Taking into account the effect of non-ideal reconciliation efficiency and non-optimal distribution of the displacement parameter discussed in the previous section, this means that, in the weak noise regime, protocols based on discretized modulations may admit the key generation only when the signal magnitude is within the certain range. A detailed investigation of the critical strength requires more refined approach and will be presented elsewhere.

On the contrary, in the strong noise regime, which is of the most interest from the perspective of low-frequency implementations of QKD, the emergence of the lower threshold appears to be rather marginal effect and the security of QKD can be investigated using the weak-signal approximation.

\section{Conclusion}

In anticipation of appearance implementations of QKD protocols in the spectral domain below the mid-infrared, we have considered a general problem of QKD protocols based on displaced thermal states with a discretized distribution of the displacement parameter. We have studied specific features of such protocols distinguishing them from well-studied protocols utilizing Gaussian states. We developed a basic formalism separating the effects of the Gaussian channel and non-Gaussian modulations. With the help of this formalism, we have studied the effect of the magnitude of the quasi-classical driving field in the source of displaced quantum states.

The main important feature, specific for protocols with discretized modulations, is the impossibility to generate secret key, in the limit of strong magnitude of the quasi-classical field. The physical origin of such collapse of QKD is weak overlap of 
the density matrices of individual states, which makes the transmitted sequence of quantum states essentially classical. In this limit, information available for eavesdropper is limited only by the entropy of the distribution of the displacement parameter, which, in turn, limits from above the mutual information between legitimate communicating parties. 

Since the emergence of the classical ensemble is due to lacunae in the factual filling the complex plane by the values of the displacement parameter used for preparation of transmitted states, it becomes a limiting factor whenever the number of transmitted states is too small even if they are sampled from the Gaussian distribution. This is a manifestation of possible quantum correlations between the transmitted states and environment (eavesdropper). This indicates that the usual incorporation of the finite-length effect through introduction of the reconciliation efficiency, which takes into account only the classical component of the QKD protocol, may not be enough to estimate correctly the possible key generation rate. 

The numerical investigation of the signal strength dependence of the key generation rate revealed that two operating regimes must be distinguished: strong and weak noise. The strong noise regime is relevant when the number of thermal photons is large and is of the most importance for low-frequency QKD implementations. In this case, the security boundary is determined by the weak signal limit and we have found its explicit form. 

The weak noise regime corresponds to a small number of thermal photons. Numerical simulations showed that in this regime a low-signal threshold may appear, so that the secret key can be generated only when the signal is sufficiently strong (but not too strong because of the transition to the classical ensemble discussed above).

%


\end{document}